\documentclass[fleqn,usenatbib]{mnras}

\usepackage{newtxtext,newtxmath}

\usepackage[T1]{fontenc}

\DeclareRobustCommand{\VAN}[3]{#2}
\let\VANthebibliography\thebibliography
\def\thebibliography{\DeclareRobustCommand{\VAN}[3]{##3}\VANthebibliography}

\usepackage{graphicx}	
\usepackage{amsmath}	

\title[Prospects for the detection of VHE pulsars with LHAASO and SWGO]{Prospects for the detection of very-high-energy pulsars with LHAASO and SWGO}

\author[Hu et al.]{
Quan Hu,$^{1,2}$
Yi Zhang,$^{1,2}$
Kaikai Duan$^{1}$\thanks{E-mail: duankk@pmo.ac.cn(KKD)}
and Houdun Zeng$^{1}$\thanks{E-mail: zhd@pmo.ac.cn(HDZ)}
\\
$^{1}$Key Laboratory of Space Science and Astronomy, Purple Mountain Observatory, Chinese Academy of Sciences, Nanjing, China\\
$^{2}$University of Science and Technology of China, Hefei, China
}

\date{Accepted XXX. Received YYY; in original form ZZZ}

\pubyear{2024}

\begin{document}
\label{firstpage}
\pagerange{\pageref{firstpage}--\pageref{lastpage}}
\maketitle

\begin{abstract}
Pulsations from the Crab pulsar have been detected by the MAGIC telescopes at energies up to 1.5 TeV, and the pulsed emission from the Vela pulsar was detected by H.E.S.S., reaching tens of TeV. These discoveries, along with the proposed additional emission due to inverse Compton scattering at TeV energies, lead us to consider suitable candidates for detection with current and future extensive air show (EAS) experiments at very-high-energy (VHE; 0.1 $-$ 100 TeV) ranges.  
Leveraging energy spectrum data from pulsars as observed by Fermi and Imaging Atmospheric Cherenkov Telescopes (IACTs) and considering the sensitivities of both LHAASO and SWGO, this study evaluates their detectability and estimates the time required for their significant detection.
Our results indicate that LHAASO could detect the Crab's pulsed signal within six years, while SWGO might detect Vela's signal within one year. Observations of the most energetic Fermi pulsars with EAS experiments will provide insight into the nature of VHE pulsar emissions, helping to clarify the primary characteristics of VHE pulsars.
\end{abstract}




\section{Introduction}

Rapidly rotating high-magnetic neutron stars, known as pulsars, emit pulsed electromagnetic radiation with periods ranging from $\sim$1 millisecond to $\sim$10 seconds. This pulsed emission spans a broad electromagnetic spectrum, from radio waves to extremely energetic multi-TeV gamma rays \citep{1997A&A...326..682B,2005AJ....129.1993M,2011AdSpR..47.1281M,2013ApJS..208...17A,2016AA...585A.133A}.Pulsed radiation has been crucial in studying pulsars' emission mechanisms and magnetosphere structures. 
Prior to the launch of Fermi-LAT, pulsar models such as the outer gap (OG)  \citep[e.g][]{2000ApJ...537..964C,2008ApJ...676..562T} and slot gap (SG) \citep[e.g][]{2008ApJ...680.1378H} models predicted that the highest-energy photons are emitted through curvature radiation, exhibiting an exponential cutoff around several GeV in the spectrum. Some refined models have extended this cutoff up to 100 GeV.
Several models propose the presence of a secondary component in pulsar emission via inverse Compton scattering (ICS) \citep[e.g.][]{1986ApJ...300..500C,1986ApJ...300..522C,1996ApJ...470..469R,2001ApJ...549..495H,2006MNRAS.366.1310T,2021ApJ...923..194H}. A more comprehensive model of the pulsar's very-high-energy (VHE) emission has been established, as indicated by \citet{2021ApJ...923..194H}. The synchro-curvature (SC) radiation can reach up to approximately 100 GeV, while the synchrotron self-Compton (SSC) radiation can extend to a few TeV, and the ICS radiation can span over 10 TeV.

The Fermi Gamma-ray Space Telescope has recently released the Third Fermi-LAT Catalog of Gamma-ray Pulsars (3PC), which incorporates data collected over up to 14 years \citep{2023ApJ...958..191S}. This catalog presents 294 pulsars with energy spectra extending beyond the GeV range. Notably, this release includes more pulsars whose spectra extend beyond 50 GeV compared to the first and second pulsar catalogs.
In addition to space-based observations, ground-based observatories have also made significant contribution to the investigation of pulsars in the VHE range. For example, pulsations from the Crab pulsar have been detected by the Major Atmospheric Gamma Imaging Cherenkov (MAGIC) telescopes at energies extending from 25 GeV to 1.5 TeV \citep{2008Sci...322.1221A,2011ApJ...742...43A,2012A&A...540A..69A,2016AA...585A.133A}, and by Very Energetic Radiation Imaging Telescope Array System (VERITAS) at energies up to about 400 GeV \citep{2011Sci...334...69V, Nguyen:2016fN}. Furthermore, MAGIC also detected pulsed emission from the Geminga pulsar in the energy range of 15 GeV to 75 GeV \citep{2016AA...591A.138A}, while High Energy Stereoscopic System (H.E.S.S.) observed the pulsed emission from the Vela pulsar in the sub-20 GeV to 100 GeV range \citep{2018A&A...620A..66H}, and from PSR B1706-44 in the sub-100 GeV energy range \citep{Spir-Jacob:2019XY}. Notably, in 2023, H.E.S.S. team published the results of the observations of the Vela pulsar, reporting the detection of the significant pulsed emission which reaches tens of TeV \citep{2023NatAs.tmp..208H}.
In contrast to the Crab pulsar, where the VHE emission may be linked to the GeV emission, the VHE gamma-ray emission from Vela pulsar is likely derived from extra mechanisms, such as inverse Compton scattering.

Investigating the potential for detecting pulsars at VHE using next-generation ground-based experiments is of great value \citep{2013APh....43..287D}. These experiments encompass both Imaging Air Cherenkov Telescope (IACT) and extensive air show (EAS) experiments. In a study of \citet{2017MNRAS.471..431B}, assuming no high-energy (HE; 0.1 $-$ 100 GeV) cutoff in the spectra of 12 Fermi pulsars with emissions above 25 GeV, 50 hours of observations were simulated using Cherenkov Telescope Array (CTA, \citealt{2019scta.book.....C}) to ascertain the detectability of such pulsars. The findings suggest that 5 $-$ 8 gamma-ray pulsars can be detected at the VHE range, with five being sufficiently bright for significant VHE detection above 0.25 TeV.

Typical large EAS experiments such as Large High Altitude Air Shower Observatory (LHAASO,  \citealp{2022ChPhC..46c0001M}) and the proposed Southern Wide-Field Gamma-ray Observatory (SWGO, \citealp{2021AN....342..431B}) cover the northern and southern hemispheres, respectively. LHAASO, a large hybrid EAS array located on Haizi Mountain in Daocheng, Sichuan province, China, consists of three sub-arrays: the one square-km array (KM2A), Water Cherenkov Detector Array (WCDA), and the wide-field air Cherenkov/fluorescence telescopes array (WFCTA). The LHAASO is a survey instrument sensitive to gamma rays with energies ranging from 100 GeV to 1~PeV, achieving the sensitivity better than 0.1 Crab nebula flux (The flux of the Crab Nebula is $8.2\times10^{-14}~\mathrm{cm}^{-2}\mathrm{s}^{-1}\mathrm{TeV}^{-1}$ at 10 TeV. Detailed spectral parameters can be found in Table~\ref{Tab1}.). Consequently, the LHAASO holds an advantage and significant potential for detecting pulsars with VHE emissions. Further details about the LHAASO can be found in \citet{2022ChPhC..46c0001M}. For sources in the southern hemisphere, SWGO is a prospective next-generation instrument with sensitivity in the VHE band.

In this study, we seek to estimate the observation time required for LHAASO and SWGO to achieve a significant detection (S $=$ 5) at energies above 1 TeV. Section 2 provides a succinct description of our estimation method. The estimated results for several pulsars, namely the Crab pulsar, Geminga pulsar, Vela pulsar, and PSR B1706-44, are detailed in Section 3. Subsequently, our discussion and conclusion are presented in Section 4.

\section{Methods}

In the phaseograms of gamma-ray pulsars, we identify the on-pulse and off-pulse intervals. For simplification in our work, we have excluded the bridge interval. To evaluate the detector's sensitivity to on-pulse signals, the significance S can be simplified when the background count significantly exceeds the signal count according to Li \& Ma's method \citep{1983ApJ...272..317L} as:
\begin{equation}
S=\frac{N^{\mathrm{onP}}_p}{\sqrt{N^{\mathrm{onP}}_{\mathrm{cr}} + N^{\mathrm{onP}}_{\mathrm{nebula}}}},
\label{eq1}
\end{equation}
where $N^{\mathrm{onP}}_p, N^{\mathrm{onP}}_{\mathrm{cr}},$ and $N^{\mathrm{onP}}_{\mathrm{nebula}}$ represent the number of pulse signal events in the on-pulse region, the cosmic ray (CR) background events, and the gamma-ray events from the external nebula, respectively.

In ground-based EAS experiments, such as LHAASO and SWGO, the angular resolution in the threshold energy range is approximately $0.5^\circ$  \citep{2022ChPhC..46c0001M}, which is significantly larger than the $0.01^\circ$ achieved by IACTs \citep{2013APh....43..171B}. As a result, a relatively large Region of Interest (ROI) is generally required to enhance detection significance. Additionally, in the multi-TeV energy range, the survival fraction of CR reaches a level of one-thousandth when performing CR/$\gamma$  discrimination in EAS experiments \citep{2022ChPhC..46c0001M}.

Therefore, we introduce $\beta$ as the ratio of the nebula to the cosmic ray background events after CR/$\gamma$ discrimination within the ROI, defined by the equation:
\begin{equation}
\beta = \frac{N_{\mathrm{nebula}}}{N_{\mathrm{cr}}} = \frac{F_{\mathrm{nebula}} r_{\gamma} f_{\rm ROI}}{F_{\mathrm{cr}} r_{\mathrm{cr}} \Delta \Omega},
\label{eq_beta}
\end{equation}
where, $F_{\mathrm{nebula}}$ and $F_{\mathrm{cr}}$ represent the fluxes of the nebula and cosmic rays, respectively; $r_{\gamma}$ and $r_{\mathrm{cr}}$ denote the survival fractions of gamma rays and cosmic rays after CR/$\gamma$ discrimination, respectively;
$\Delta \Omega$ is the solid angle of the region of interest (ROI). The radius of the ROI is defined as 1.51 times the Gaussian width of the point spread function (PSF), which encompasses 68\% of the signals from point sources \citep{2004HNP..28..1094}. $f_{\rm ROI}$ represents the fraction of events from the nebula within the ROI.
Thus, the significance formula can be simplified to
\begin{equation}
S = \frac{N^{\mathrm{onP}}_p}{\sqrt{(1 + \beta) N^{\mathrm{onP}}_{\mathrm{cr}}}},
\label{eq2}
\end{equation}

For the LHAASO and SWGO detectors, the differential sensitivity, $\mathrm{d}F_s$, is defined as:
\begin{equation}
    \mathrm{d}F_s = \frac{\mathrm{d}N_s}{\mathrm{d}A\mathrm{d}t\mathrm{d}E},
    \label{eq3}
\end{equation}
where $F_s$ reflects the detectors' sensitivity after one year of operation, and $N_s = 5\sqrt{N_{\mathrm{cr}}}$ represents the minimum number of detectable events from a specific direction that achieves a significance level of $5$ within the same exposure time.
In this context, $N_{\mathrm{cr}}$ is the number of background events in a particular direction, $\mathrm{d}A$ denotes the effective detection area for the energy bin, $\mathrm{d}t$ is the total observation time, and $\mathrm{d}E$ is the energy bin width.

Given that a pulsar and its nebula are essentially co-located, sharing the same differential area ($\mathrm{d}A$) and observation time ($\mathrm{d}t$), we can deduce the following relationship:
\begin{equation}
N^{\mathrm{onP}}_p = N_s \frac{F^{\mathrm{onP}}}{F_s}.
\label{eq4}
\end{equation}

Letting \(\alpha\) represent the on-pulse fraction, we obtain:
\begin{equation}
N^{\mathrm{onP}}_{\mathrm{cr}} = \alpha  N_{\mathrm{cr}}.
\label{eq5}
\end{equation}

Substituting the aforementioned equations into equation \ref{eq2}, the annual significance, $S$, is determined by:

\begin{equation}
S = \frac{5}{\sqrt{\alpha (1 + \beta)}} \frac{F^{\mathrm{onP}}}{F_s}.
\label{eq6}
\end{equation}

Typically, a significance threshold of \(S=5\) is adopted for the sensitivity curve. Therefore, the sensitivity, $F_s^{\mathrm{onP}}$, for observing pulsed emissions over $n$ years for a source at a consistent declination is expressed as:

\begin{equation}
F_s^{\mathrm{onP}} = F_s \sqrt{\frac{\alpha (1 + \beta)}{n}}.
\label{eq7}
\end{equation}

\section{Results for Crab, Geminga, Vela pulsars and PSR B1706-44}

We implement a systematic selection process to select pulsars exhibiting both observed and potential TeV radiations. Based on the Third Fermi-LAT Catalog of Gamma-ray Pulsars, we initially chose pulsars with pulsed radiation above 50 GeV as candidates. Further scrutiny, using the TeVCatalog\footnote{http://tevcat2.uchicago.edu/}, lead us to select four pulsars— the Crab pulsar, the Geminga pulsar, the Vela pulsar, and PSR B1706-44 —as likely sources of pulsed TeV emissions. We employ this methodology to evaluate the observational capabilities of LHAASO and SWGO for these pulsars.

\begin{figure*}
    \centering
    \includegraphics[width=0.45\textwidth]{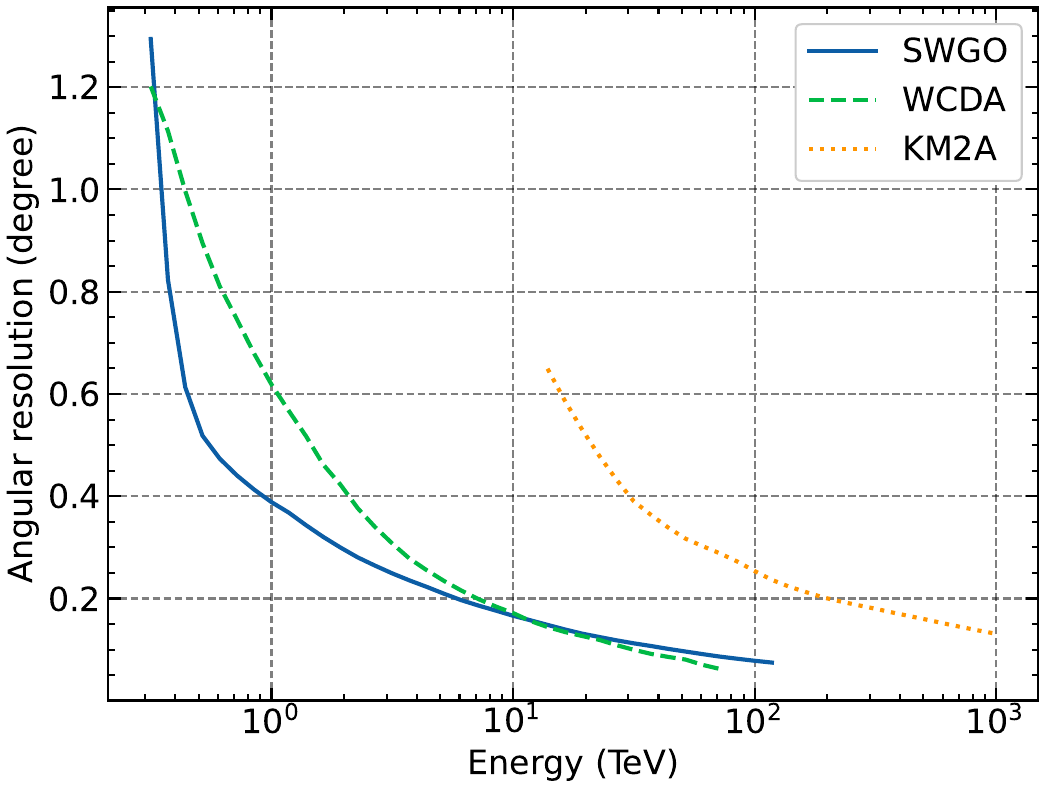}
    \includegraphics[width=0.45\textwidth]{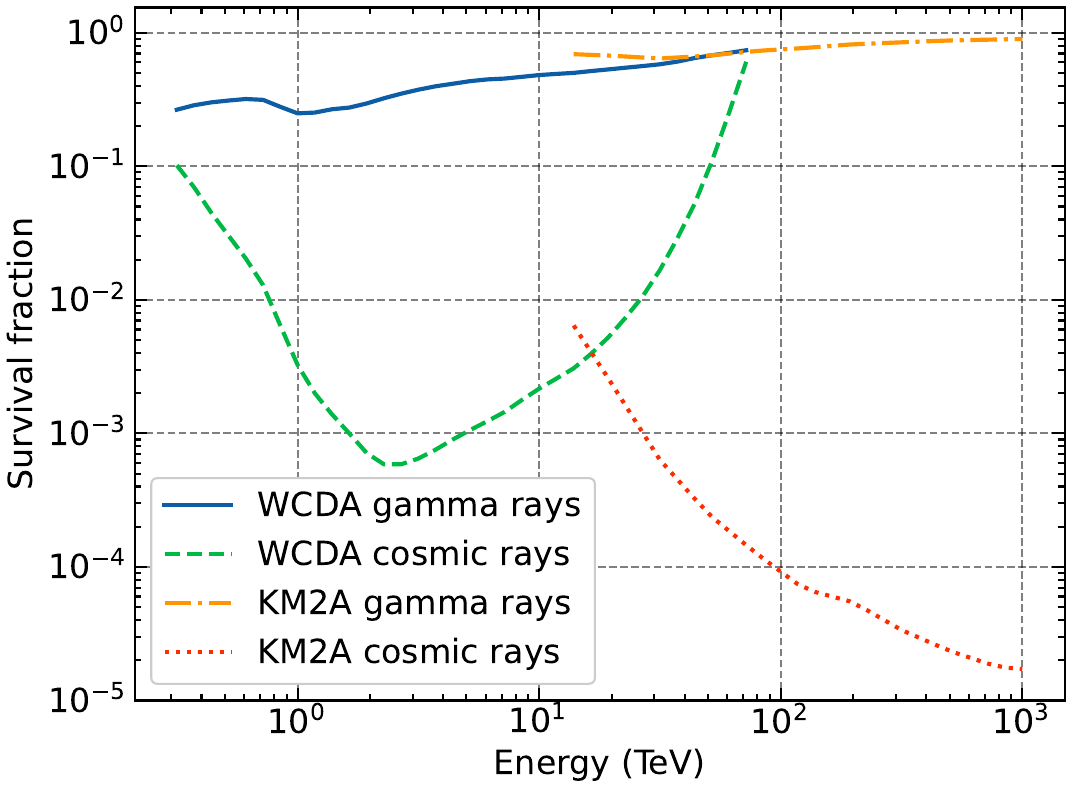}
    \includegraphics[width=0.45\textwidth]{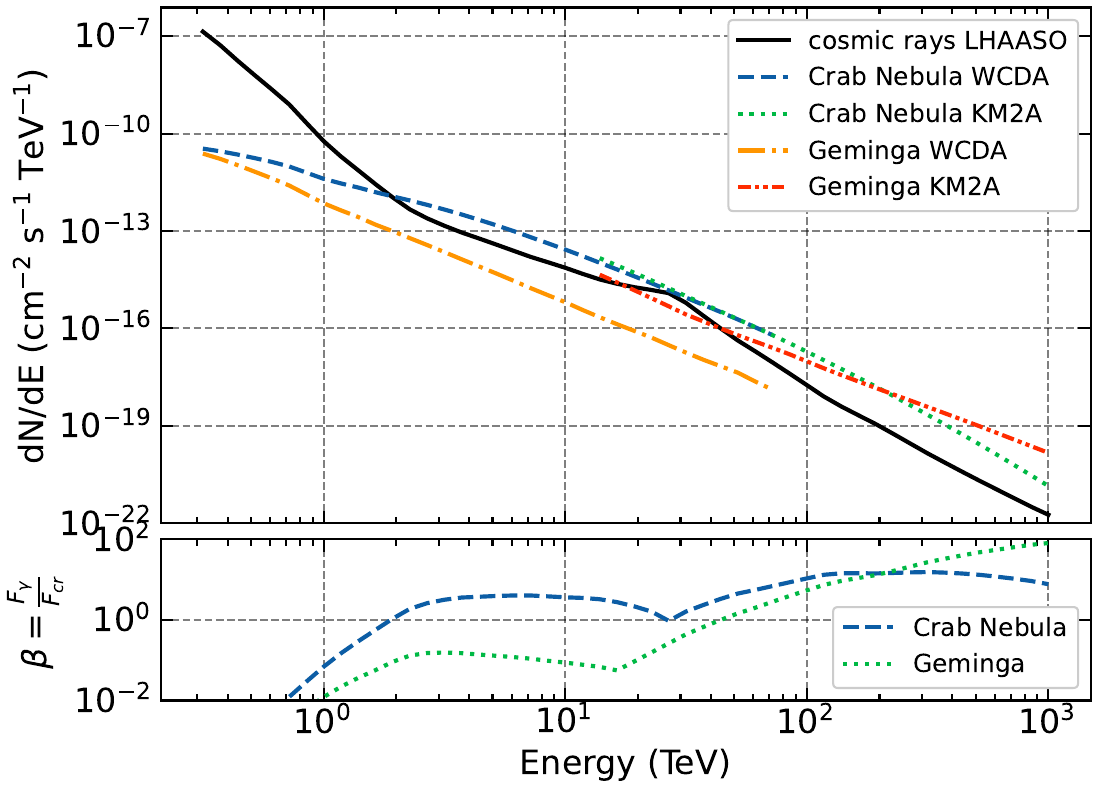}
    \includegraphics[width=0.45\textwidth]{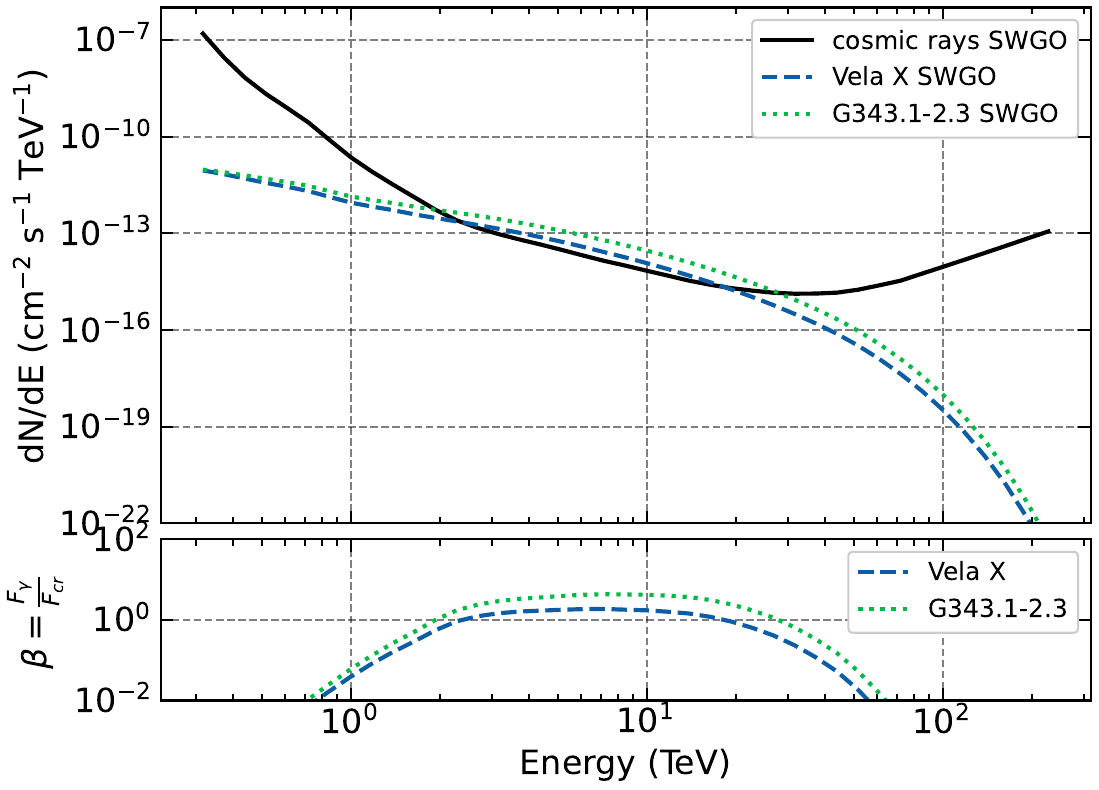}
    \caption{The angular resolutions of LHAASO-WCDA, LHAASO-KM2A and SWGO (upper left) and the survival fractions of WCDA and KM2A for gamma rays and cosmic rays (upper right), which are taken from \citet{2022ChPhC..46c0001M}, \citet{2024arXiv240101038L} and \citet{2019arXiv190208429A}. The survival fraction of SWGO is assumed to be same to that of WCDA. The flux of cosmic rays and pulsar nebulae within the solid angle $\Delta \Omega$ after  CR/$\gamma$ discrimination, and the energy-dependent ratio of $\beta$ for Crab Nebula and Geminga observed by LHAASO (lower left) and for Vela X and G343.1-2.3 observed by SWGO (lower right). The radius of ROI is set at 1.51 times angular resolution, which encompasses 68\% of the signals from point sources.}
    \label{fig:J_cr_gamma}
\end{figure*}


Table~\ref{Tab1} displays the surrounding pulsar wind nebulae and the extensions and spectral parameters of these nebulae for the four VHE) pulsars, as obtained from observations of the respective experiments.

With LHAASO situated at approximately 30 degrees north latitude \citep{2022ChPhC..46c0001M}, both the Crab and Geminga pulsars fall within its field of view, with minimum zenith angles of less than 30$^{\circ}$. Additionally, SWGO is anticipated to be situated near a latitude of 23$^{\circ}$S \citep{2023JPhCS2429a2022L}, providing coverage of the directions of Vela pulsar and PSR B1706-44.

\begin{table*}
  \centering
  \caption{The extensions and spectral parameters of Pulsar Wind Nebulae (PWN) surrounding four very-high-energy pulsars, as observed by IACTs.}
  \resizebox{\textwidth}{!}{
  
  
  \begin{tabular}{llccccc}
    \hline
    Pulsar         & surrounding PWN   & $\sigma$ (deg) &  $N_0$ ($\mathrm{cm}^{-2}\mathrm{s}^{-1}\mathrm{TeV}^{-1}$)  & $E_0$ (TeV) & $\Gamma$ & $E_{\rm cut}$ (TeV) \\
    \hline
    Crab pulsar    & Crab Nebula   & 0.014 &  $(8.2 \pm 0.2) \times 10^{-14}$      & 10 & $(2.09 \pm 0.01) + (0.19 \pm 0.02)\log_{10}(E/E_0)$ & - \\%
    Geminga pulsar & Geminga       & 1.3 &  $13.6^{+2.0}_{-1.7} \times 10^{-15}$ & 20 & $2.34 \pm 0.07$ & - \\%
    Vela pulsar    & Vela X     & 0.51 &  $(11.6 \pm 0.6) \times 10^{-12}$     & 1  & $1.36 \pm 0.06$ & $13.9 \pm 1.6$ \\%
    PSR B1706-44   & G343.1-2.3 & 0.29 &  $(4.2 \pm 0.8) \times 10^{-12}$      & 1  & $2.0 \pm 0.1$   & - \\%
    \hline
  \end{tabular}
  }
  Note. The spatial distributions of PWNs are characterized by a symmetric 2D Gaussian distribution $\frac{1}{\sigma^2} e^{-\frac{x^2}{2\sigma^2}}$ with an extension of $\sigma$. References for the extensions and spectral parameters: Crab Nebula: \cite{2020NatAs...4..167H} and \cite{2021Sci...373..425L}; Geminga: \cite{2009ApJ...700L.127A} and \cite{2017Sci...358..911A}; Vela X: \cite{2012AandA...548A..38A}; G343.1-2.3: \cite{2011AandA...528A.143H}.
  \label{Tab1}
\end{table*}

\begin{figure*}
    \centering
    \includegraphics[width=0.4\textwidth]{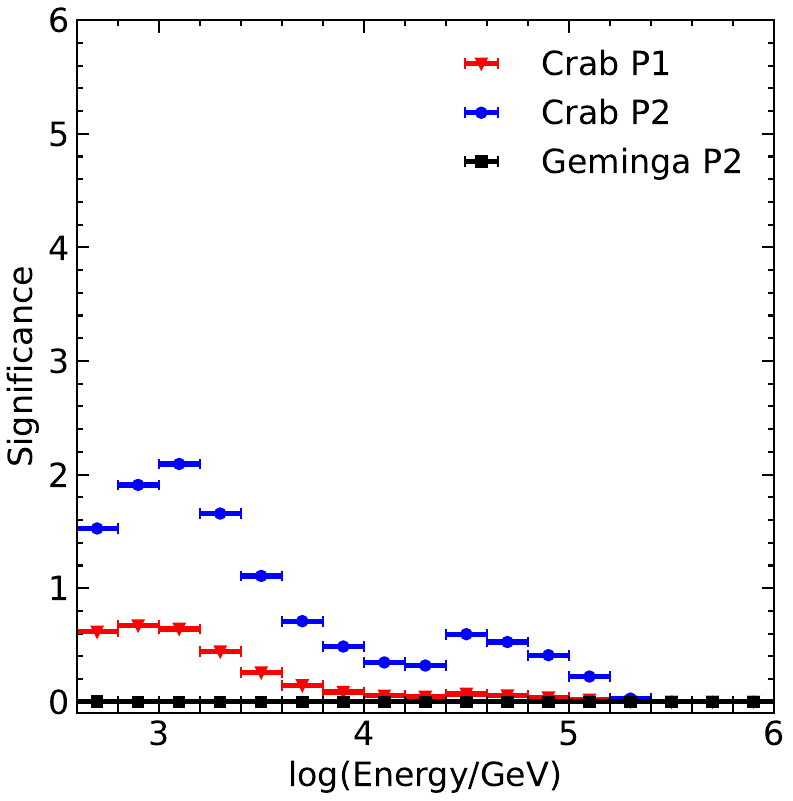}
    \includegraphics[width=0.42\textwidth]{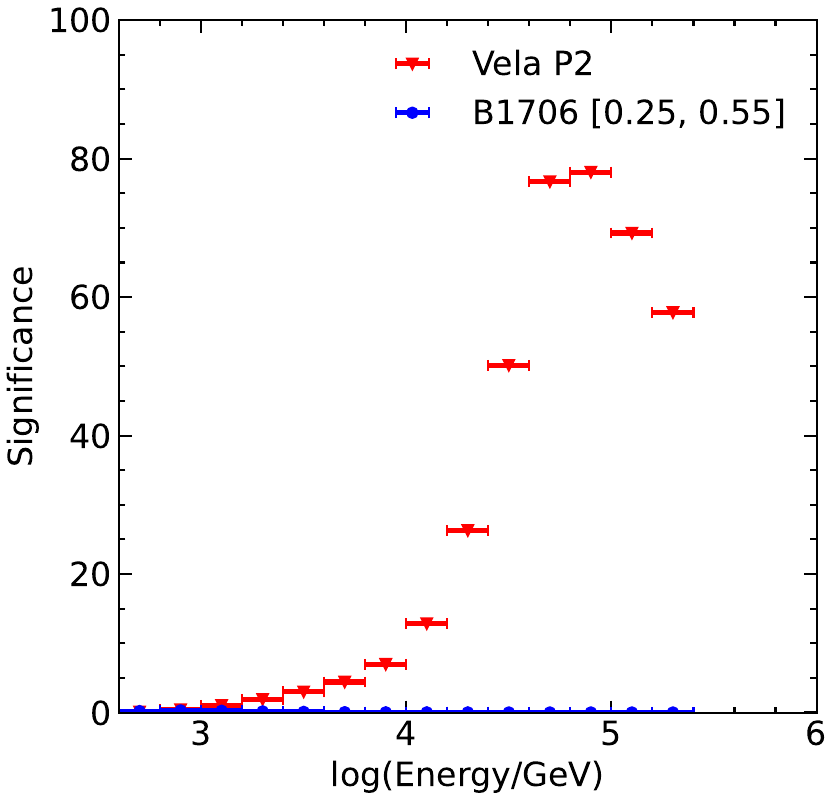}
    \caption{Expected significance of on-pulse emission from the Crab and Geminga pulsars over a one-year observation period by LHAASO (left), and from the Vela and B1706-44 pulsars over a one-year observation period by SWGO (right).}
    \label{Fig1}
\end{figure*}

Based on the gamma-ray flux of each pulsar wind nebula, the CR flux, and the instrument's CR/$\gamma$ discrimination capability, we calculate the ratio $\beta$ using Equation~\ref{eq_beta}.
The flux of the nebulae in the TeV energy range, derived from specific experimental observations, follows distinct spectral distributions. The energy spectrum of the Crab Nebula follows a log-parabolic function \citep{2021Sci...373..425L}, while Vela X follows an exponentially cutoff power-law distribution \citep{2012AandA...548A..38A}. In contrast, the spectra of Geminga and G343.1-2.3 are described by a power-law \citep{2017Sci...358..911A,2011AandA...528A.143H}. The parameters of the SEDs are listed in Table 1.

The angular resolution and survival fractions of gamma rays and cosmic rays after CR/$\gamma$ discrimination for LHAASO-WCDA and LHAASO-KM2A are obtained from \citet{2022ChPhC..46c0001M} and \citet{2024arXiv240101038L}, respectively. The angular resolution of SWGO is derived from \citet{2019arXiv190208429A}, and its survival fraction is assumed to be similar to that of LHAASO-WCDA. 
The values of angular resolution, survival fractions of gamma rays and cosmic rays are shown in Figure \ref{fig:J_cr_gamma}. The radius of the ROI is energy-dependent and is set at 1.51 times the Gaussian width of the angular resolution. For simplicity, we used the same ROI for pulsars observed with the same instrument. We assume that the pulsar wind nebulae are located at the same positions as the pulsars and that their spatial distributions can be characterized by a symmetric 2-dimensional (2D) Gaussian distribution. We then estimate the 2D gamma-ray distribution of the pulsar wind nebulae as the convolution of their Gaussian distribution with the PSF of the instrument. Using this 2D gamma-ray distribution, we calculate the fraction of gamma rays for each nebula within the ROI.

The cosmic ray flux is derived from the cosmic ray model \citep{2019arXiv191003721S}. 
The flux and the ratio of the nebula to cosmic ray background after CR/$\gamma$ discrimination are depicted in Figure~\ref{fig:J_cr_gamma}. From one to $\sim$ 100 TeV, $\beta$ exceeds 1, indicating the importance of considering the gamma-ray background from the nebula in this energy range.

We extrapolated the flux of pulsar emissions to VHE by employing the directly observed spectral energy distributions (SEDs) of the Crab and Vela pulsars, and the GeV energy band observations of the Geminga pulsar and PSR B1706-44, utilizing a power-law spectral model. The parameters of this power-law spectral model are listed in Table \ref{Tab2}. For instance, the energy spectra of the Crab pulsar's components P1 (main pulse phase from $-0.017$ to 0.026) and P2 (interpulse phases from 0.377 to 0.422) have been measured up to 600 GeV and 1.5 TeV, respectively, displaying power-law spectral indices of 3.2 $\pm$ 0.4 and 2.9 $\pm$ 0.2 \citep{2016AA...585A.133A}. The energy spectrum of the Geminga pulsar's P2 (the second pulse seen by Fermi-LAT, with pulse phase from 0.550 to 0.642), ranging from 10 GeV to 75 GeV with a power-law spetral index of 5.62 $\pm$ 0.54 \citep{2020AandA...643L..14M}, and the energy spectrum of PSR B1706-44, (within the on-pulse range observed by H.E.S.S, with pulse phase from 0.25 to 0.55), with a power-law spetral index of 3.76 $\pm$ 0.36 \citep{Spir-Jacob:2019XY} are both extrapolated to the TeV range. Recent results from \citet{2023NatAs.tmp..208H} indicates that the Vela pulsar's P2 component (with pulse phase from 0.55 to 0.60), spanning from 260 GeV to 28.5 TeV, possesses a hard power-law spectral index of 1.4 $\pm$ 0.3. 
The differential sensitivity of LHAASO is described in \cite{2016NPPP..279..166D}, while that of SWGO is derived from simulations reported in  \cite{2019BAAS...51g.109H}.

\begin{table*}
  \centering
  \caption{Spectral parameters and on-pulse fraction of pulsar emissions and the time needed to achieve a five-sigma detection.}
  \begin{tabular}{lcccccc}
    \hline
    Pulsar phase              & $E_0 (\mathrm{GeV})$    & $N_0 (\mathrm{TeV}^{-1}\mathrm{cm}^{-2}\mathrm{s}^{-1})$  & $\Gamma$ & On-pulse fraction & Experiment     & Expected time (year)\\
    \hline
    Crab P1 [-0.017–0.026]          & 150     & $(1.1 \pm 0.3)\times10^{-11}$   & 3.2 $\pm$ 0.4  & 0.043 & LHAASO    & 55 at 630~GeV\\
    Crab P2 [0.377–0.422]          & 150     & $(2.0 \pm 0.3)\times10^{-11}$   & 2.9 $\pm$ 0.2  & 0.045 & LHAASO    & 6 at 1~TeV\\
    Geminga P2 [0.550–0.642]        & 32.15   & $(2.28 \pm 0.74) \times10^{-9}$   & 5.62 $\pm$ 0.54 & 0.092 & LHAASO   & >1000\\
    Vela P2 [0.55–0.6]          & 4240    & $(1.74 \pm 0.52) \times10^{-15}$  & 1.4 $\pm$ 0.3  & 0.05  & SWGO      & $<$1 at $>$6.3~TeV\\
    B1706-44 [0.25-0.55]  & 20      & $(4.3 \pm 0.9) \times10^{-8}$    & 3.76 $\pm$ 0.36 & 0.3   & SWGO     & 287 at 630~GeV\\
    \hline
  \end{tabular}\\
   Note. The differential energy spectra are modeled with a power law form with $N_0(E/E_0)^{-\Gamma}$. Errors indicate 1$\sigma$ statistical uncertainties. \\ References for the spectral parameters: Crab P1: \cite{2016AA...585A.133A};  
   Crab P2: \cite{2016AA...585A.133A};
   Geminga P2:\\ \cite{2020AandA...643L..14M};
   Vela P2: \cite{2023NatAs.tmp..208H};
   B1706-44: \cite{Spir-Jacob:2019XY}.
  \label{Tab2}
\end{table*}

\begin{figure*}
  \centering
  \includegraphics[width=0.45\textwidth]{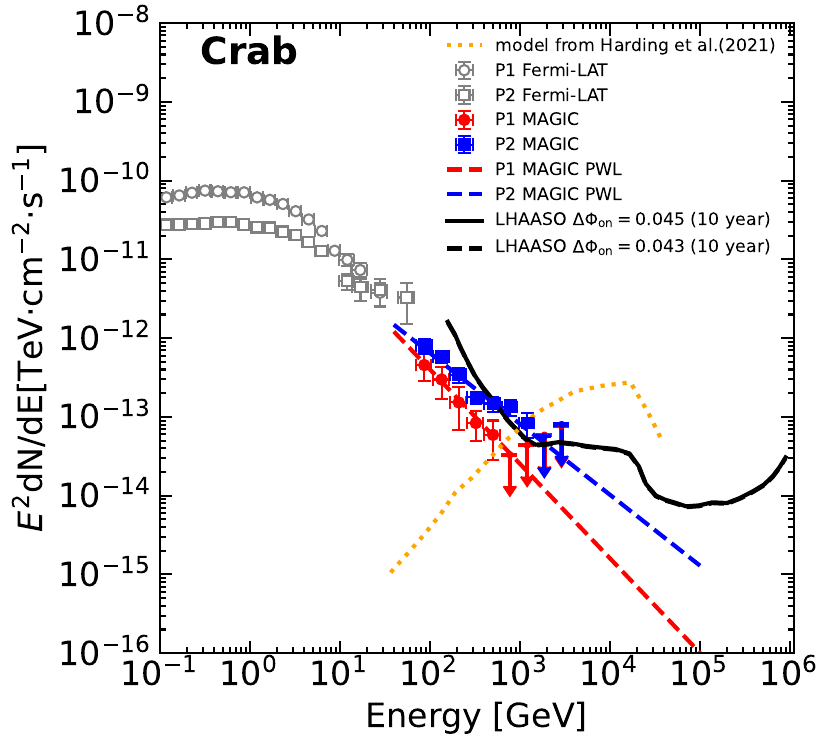}
  \includegraphics[width=0.45\textwidth]{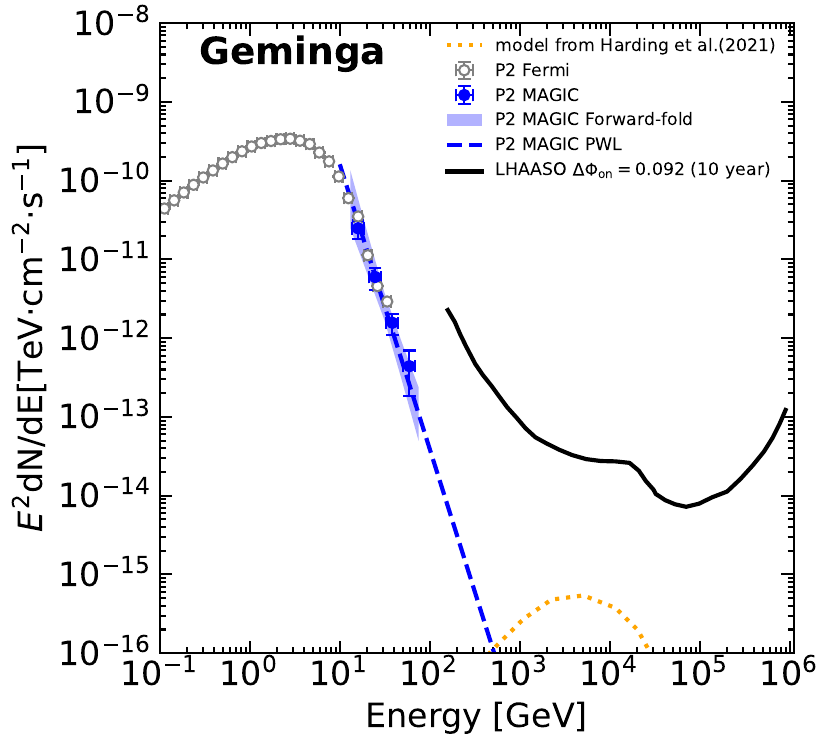}
  \includegraphics[width=0.45\textwidth]{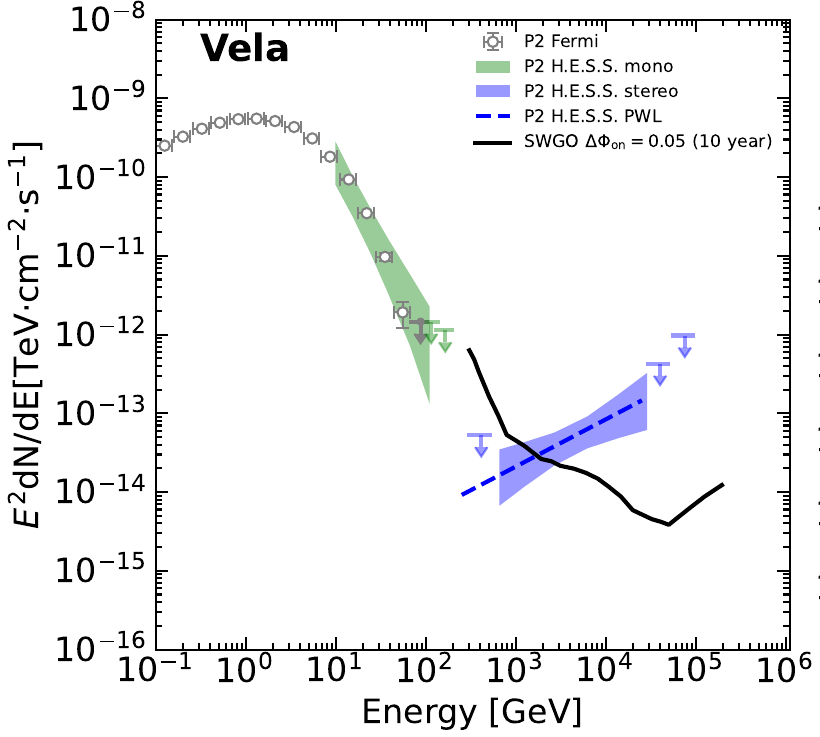}
  \includegraphics[width=0.45\textwidth]{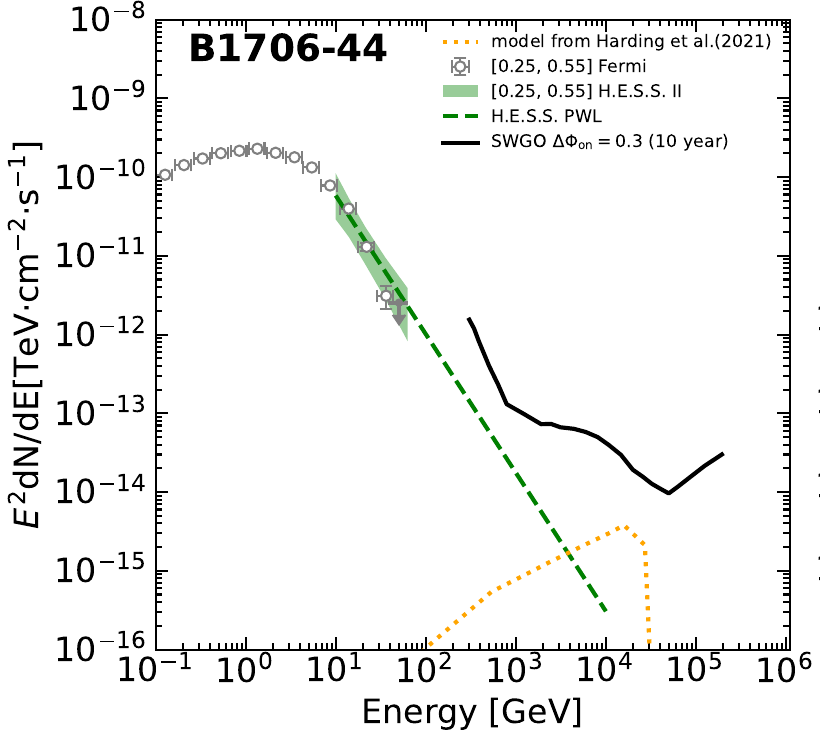}
  \caption{Energy spectra of on-pulse emissions from the Crab pulsar (upper left), Geminga pulsar (upper right), Vela pulsar (bottom left), and B1706-44 (bottom right), compared to the ten-year sensitivity of LHAASO and SWGO. The orange dotted lines indicate the predicted VHE flux from \citet{2021ApJ...923..194H}. Grey data points correspond to observations by Fermi from the following sources: Crab (\citealt{2014A&A...565L..12A}), Geminga (\citealt{2020AandA...643L..14M}), Vela (\citealt{2018A&A...620A..66H}), and B1706-44 (\citealt{Spir-Jacob:2019XY}). Details of the colored data or butterfly area and associated fitted dashed lines are provided in Table \ref{Tab2}.
  }
  \label{Fig2}
\end{figure*}

We calculated the significance of pulsed emissions in each energy bin after one year's observation using equation \ref{eq6}, and the results are depicted in Figure~\ref{Fig1}. In LHAASO's observations, the pulsed emission from the Crab pulsar shows the highest significance near 1 TeV, while the emission from the Geminga pulsar is rarely detected in the VHE range in a short time span of 1 year. In contrast, in SWGO's observations, the Vela pulsar demonstrates the highest significance at multi-TeV, and the observation of PSR B1706-44 achieves the highest significance at $\sim$630 GeV.
The time required for LHAASO and SWGO to observe these pulsars with five-sigma significance is listed in the last column of Table 2. We used both LHAASO-WCDA and LHAASO-KM2A to calculate this time because their sensitivity energy ranges overlap and complement each other. The minimum time needed for the above sources falls within the WCDA energy range. 
Based on estimates, LHAASO-WCDA is expected to require less than $\sim$7 years to observe the pulsed emission from the Crab pulsar in the VHE range, while SWGO could observe Vela's pulsed emission within one year. In Figure~\ref{Fig2}, the energy spectra of these four pulsars are depicted. The grey data points represent measurements from Fermi's experiment, while the blue, red, and green data points correspond to data from MAGIC and H.E.S.S. experiments, with the respective colored dashed lines representing the power-law spectra. The solid black line illustrates the detection sensitivity of LHAASO and SWGO experiments for different on-pulse fractions of pulsed radiation over a 10-year observation period. The orange dotted line represents the VHE spectra of sources such as Crab and Geminga pulsars, based on the predictions of \citet{2021ApJ...923..194H}.

\section{Conclusions and Discussion}
We presented the prospects of detecting four gamma-ray pulsars (Crab, Geminga, Vela Pulsar and PSR B1706-44) in the VHE range using LHAASO and SWGO. The required observation time to achieve the necessary statistical significance was calculated for the four potential TeV pulsar sources, assuming that the pulsed emission has no HE cutoff and that the energy spectrum can be described by a single power law. This prospect is estimated based on the detectors' sensitivity and performance for stable gamma-ray sources. In the most optimistic scenario, LHAASO-WCDA is expected to require less than $\sim$6 years to observe the pulsed emission from the Crab pulsar in the VHE range, while Vela's pulsed emission can be observed by SWGO in one year.

The energy spectra used in the analysis of this paper are the results obtained by fitting the data from H.E.S.S., Fermi, and MAGIC detectors with a single power law (PWL). However, the presence of a spectral cutoff due to the limit of particle acceleration capabilities is more common in the high-energy regime \citep[e.g][]{2008ApJ...676..562T,2021ApJ...923..194H}.
For the Crab pulsar, the model proposed by \citet{2021ApJ...923..194H} predicts VHE components with a cutoff, as illustrated in Figure~\ref{Fig2}. These components are significantly more pronounced than the power-law extended component at E> 1 TeV. Such components are expected to be detectable by LHAASO within a few years. Conversely, in the absence of an ICS contribution, LHAASO will be capable of detecting the Crab pulsar exhibiting a PWL spectrum up to a certain energy ( $\sim$1 TeV).
However, for Geminga pulsar and B1706-44, their predicted flux may be challenging to be observed by LHAASO and SWGO within 10 years, as illustrated in Figure~\ref{Fig2}.

As for Vela pulsar, H.E.S.S. has detected about a dozen events above 5 TeV  from it's direction \citep{2023NatAs.tmp..208H}. In the 260 GeV to 28.5 TeV energy range, the PWL spectrum is very hard with a normalization energy $E_0 = 4.24$ TeV, photon index $\alpha = 1.4$, and normalization flux A$ = 1.74\times10^{-15}$ erg$^{-1}$cm$^{-2}$s$^{-1}$. In this energy range, it is possible for SWGO to detect Vela's pulsed emission within one year.

It is worth mentioning that the fraction of the on-pulse interval, $\alpha$, has been shown to decrease with increasing energy \citep{2011ApJ...742...43A}. This makes the on-pulse emission from the pulsar more easily detectable in the very-high-energy band, as the number of background events is directly proportional to $\alpha$, while the flux remains constant.

To date, the Vela pulsar is the only one identified with additional emission components in the multi-TeV range by IACTs at E>1 TeV \citep{2023NatAs.tmp..208H}. EAS experiments would be essential in their monitoring if other pulsars exhibit similar additional emission components in the multi-TeV range.

EAS experiments such as LHAASO and SWGO have the advantages of high duty cycles and large field of view, enabling continuous monitoring of pulsars at very-high-energy. With the current sensitivity, they will allow us to study the annual-scale spectral variations of the pulsars which emit in the TeV energy range. Observations of these pulsars using EAS experiments at energies above 1 TeV will help to unravel the mechanism of the TeV component discovered in the Crab and Vela pulsars, and to clarify which class of gamma-ray pulsars have this component.

\section*{Acknowledgements}

This work is also supported by the National Natural Science Foundation of China under grants 12273114 and 12333006, the Project for Young Scientists in Basic Research of Chinese Academy of Sciences under grant YSBR-061 and the Program for Innovative Talents and Entrepreneur in Jiangsu.

\section*{Data Availability}

 All results in this paper are obtained using publicly available data.
 



\bibliographystyle{mnras}
\bibliography{ms} 








\bsp	
\label{lastpage}
\end{document}